\documentclass[nofootinbib,aps,prl,a4paper,twocolumn,english,superscriptaddress,longbibliography,reprint]{revtex4-2}
\usepackage[mathlines]{lineno}
\usepackage{amsthm}
\usepackage{amsmath}
\usepackage{amssymb}
\usepackage{graphicx}
\usepackage{bm}
\usepackage{color}
\usepackage[dvipsnames]{xcolor}
\usepackage{enumitem}
\usepackage{mathrsfs}
\usepackage{verbatim}
\usepackage{bbold}
\usepackage{braket}
\usepackage{lipsum}
\usepackage{url}
\usepackage{tikz}
\usepackage{graphicx}
\usepackage[colorlinks=true,citecolor=blue,urlcolor=blue]{hyperref}

\newcommand{\tr}{\text{Tr}}
\newcommand{\ave}[1]{\langle #1 \rangle}
\newcommand{\cl}[1]{\hat{\mathcal{#1}}}

\newcommand{\new}[1]{\textcolor{black}{#1}}

\usepackage[normalem]{ulem}

\begin{document}   
\title{Conserved quantities enable the quantum Mpemba effect in weakly open systems}
\author{Iris Ul\v{c}akar}
\affiliation{Jo\v{z}ef Stefan Institute, 1000 Ljubljana, Slovenia}
\affiliation{University of Ljubljana, Faculty for physics and mathematics, 1000 Ljubljana, Slovenia}
\author{Rustem Sharipov}
\affiliation{University of Ljubljana, Faculty for physics and mathematics, 1000 Ljubljana, Slovenia}
\author{Gianluca Lagnese}
\affiliation{Jo\v{z}ef Stefan Institute, 1000 Ljubljana, Slovenia}
\author{Zala Lenar\v{c}i\v{c}}
\affiliation{Jo\v{z}ef Stefan Institute, 1000 Ljubljana, Slovenia}

\begin{abstract} 
Observation of the quantum Mpemba effect has spurred much interest in its enabling conditions and its relation to the classical counterpart. Here, we consider weakly open many-body quantum systems initialized in different thermal states and examine when the initially farther state relaxes to the (non-equilibrium) steady state faster. We claim that the number of conserved quantities in the unitary part plays a crucial role: the Mpemba effect is possible only when the Hamiltonian commutes with other extensive operators or is integrable. 
The reason lies in the dynamical evolution happening in spaces of different dimensions. 
When energy is the only approximately conserved quantity, dissipation pushes the dynamics within a single-parameter manifold of different thermal states. In contrast, for Hamiltonians with several conserved quantities, the dynamics drift in the multi-dimensional space of generalized Gibbs ensembles, whose distance to the steady state is less trivial. We provide numerical results for large system sizes using tensor networks and free-fermion techniques, thereby supporting our claim.

\end{abstract}

\maketitle

The Mpemba effect is the counter-intuitive phenomenon where, under certain conditions, a hotter system cools faster than a colder one. Its occurrence challenges the conventional thermodynamic intuition and, since its popularization in the 1989~\cite{mo-69}, sparkled numerous debates and scientific investigations~\cite{glcg-11, ahn-16, lvps-17, keller-18, hu-18, bj-19, sl-22, hr-22, tyr-23, gamlmms-24, lr-17, krhv-19, kb-20, bkc-21, gr-20, kcb-22, wv-22, wbv-23, bwv-23, santos-24, vvh-24, rev} without a universally accepted explanation of the original experiment~\cite{bl-16, bh-20}. 
Recent research revealed quantum analogs of the Mpemba effect \cite{ares2025}, which can be divided into two classes: isolated and open quantum systems.

In closed quantum systems, the effect was first observed in processes of symmetry restoration in integrable Hamiltonian models. The system is prepared in different pure states that violate a symmetry of the Hamiltonian to a different degree, and the effect is naturally detected by monitoring a suitable quantification of the asymmetry during the time evolution after a quench \cite{amc-23,amvc-23,murciano2024,2rylands2024}. The phenomenon was also confirmed experimentally with trapped ions \cite{joshi-24}. Furthermore, based on the quasi-particle picture \cite{calabrese2005,alba2017,alba2018} allowed by the integrable structure, a microscopic interpretation was proposed \cite{rylands2024}.
Subsequently, the analysis has been extended to a broader class of systems such as quantum circuits \cite{klobas2024,turkeshi2024,shuo2024,klobas2025,yu2025,2ares2025,foligno2025,aditya2025}, many-body localized systems \cite{shuo2024a}, two-dimensional systems of free bosons and free fermions \cite{yamashika2024,yamashika2025}, and chaotic systems without any conserved quantities \cite{bhore2025}.

\begin{figure}
\includegraphics[width=\columnwidth]{Figure/Fig_1.pdf}
\caption{
We consider many-body quantum systems with unitary Hamiltonian and a weak dissipative evolution, initialized in two thermal states. $N_C=1$: For Hamiltonians without additional conserved quantities, each dynamics can be approximated by a thermal ensemble with a time-dependent temperature.  $N_C>1$: For Hamiltonians with more conserved quantities, dissipation can steer the evolution from thermal states in the multi-dimensional manifold of generalized Gibbs ensembles (GGEs). When projecting the dynamics onto a scalar - the distance between the time-evolved and the non-equilibrium steady state (NESS) - crossings can happen in the presence of multiple approximately conserved quantities. 
}
\label{figure:fig1}
\end{figure}

The second type of models that can exhibit the quantum Mpemba effect are open (driven) quantum systems, where the system interacts with an environment. Strategies to realize the Mpemba effect include: (i)~orthogonalization of the initial state with respect to the slowest decaying mode \cite{cll-21,goold,kcl-22,bh-22,goold2,Hayakawa0,Hayakawa1,shapira,ias-23, zhou-23, liu-24,fs-24,mcblg-24, gsm-24,westhoff2025}; 
(ii)~few-body systems of qubit(s), quantum dots, or harmonic oscillator coupled to thermal baths~\cite{goold2,Hayakawa0,Hayakawa1,shapira,ias-23, zhou-23, ww-24, liu-24,fs-24,mcblg-24, gsm-24,qww-24,befb-24, longhi-24-3,kcm-24,nava-24, ww-24,zhang2024,ew-24}; 
(iii)~a cooling protocol in a system with a metastable configuration due to a first-order phase transition \cite{nf-19} and a supercooling protocol with a strong coupling to a thermal bath \cite{wsw-24};
and (iv)~proposals for quantum-optical implementations employing different initial states of light~\cite{longhi-24,longhi-24-2}.
Some of these strategies have classical Markovian analogues, such as the orthogonalization procedure~\cite{lr-17, krhv-19,kb-20, bkc-21} and supercooling protocols~\cite{auerbach-95, ers-08}. 
Stability of symmetry restoration has also been tested with respect to the presence of dissipation~\cite{joshi-24,vitale2025}.

In this Letter, we provide a general and unifying principle for the occurrence of the Mpemba effect in thermodynamically large weakly open systems, based on the local symmetries of the Hamiltonian. In resemblance to the classical setup, we discuss systems initialized in mixed thermal states, thus not pursuing effects due to fine tuned coherence of the initial state. We propagate the density matrix with a unitary Hamiltonian evolution in the presence of weak coupling to nonequilibrium Lindblad baths and ask how the occurrence of the Mpemba effect --- a farther state relaxing to the steady state faster than a closer one --- depends on the generic properties of the many-body Hamiltonian. We claim that the number of conserved quantities of the underlying Hamiltonian, being weakly broken by the dissipation, is crucial. We give evidence that chaotic many-body models with Hamiltonian as the only approximate conserved quantity cannot exhibit the Mpemba effect, while systems with additional approximate conserved quantities --- integrable Hamiltonians being the extreme case --- allow for it. Using tensor network and free fermions approaches, we provide large system sizes $80 \geq L \geq 400$ results supporting our claim. \new{We also comment on the feasibility of experimental verification in terms of observables.}

{\it Setup.}
We consider weakly dissipative quantum many-body systems with a dominant unitary evolution. Weak dissipation is encoded in bulk Lindblad operators $L_j$ acting around every site $j$ at a rate $\epsilon \ll 1$, 
\begin{equation}
\dot{\rho} = -i [H,\rho] + \cl{D} \rho, \ 
 \cl{D} \rho = \epsilon \sum_j L_j \rho L_j^\dagger - \frac{1}{2}\{L_j^\dagger L_j, \rho\}.
 \label{eq::liouvillian}
\end{equation}
In the weak-dissipation regime, the density matrix describing the system can be decomposed as \cite{lenarcic18}
\begin{equation}\label{eq:rho}
\rho(t) = \rho_{\boldsymbol\lambda}(t) + \delta\rho(t),    
\end{equation}
where $\rho_{\boldsymbol\lambda}(t)$ is a dominant \new{$\mathcal{O}(1)$} component and $\delta\rho(t)$ a small $O(\epsilon)$ correction.
The conserved quantities $C_i$ of the unitary part, $[H,C_i]=0$, play an important role in the parametrization $\boldsymbol{\lambda}=\{\lambda_i\}_{i=1}^{N_C}$ of the zeroth order approximation $\rho_{\boldsymbol\lambda}(t)$. Namely, one can use a statistical description by introducing a Lagrange parameter $\lambda_i$ for each approximately conserved quantity \cite{lenarcic18a,lenarcic20,shirai20,lange17,lange18,reiter21,schmitt22,ulcakar23,ulcakar24,bouchoule20,rossini21,gerbino23,perfetto23,rowlands23,riggio24,starchl22,starchl24, lumia24}, 
\begin{equation}\label{eq:rho0}
\rho_{\boldsymbol\lambda}(t) = \frac{e^{-\sum_{i=1}^{N_C} \lambda_i(t) C_i}}{\tr[e^{-\sum_{i=1}^{N_C} \lambda_i(t) C_i}]}.
\end{equation}
When the Hamiltonian is the only approximate conserved quantity ($N_C=1$), $\rho_{\boldsymbol\lambda}(t)$ takes the form of a time dependent Gibbs ensemble $\rho_{\beta}(t) = e^{-\beta(t) H}/\tr[e^{-\beta(t) H}]$, parametrized with a time dependent inverse temperature
\cite{lenarcic18a,lenarcic20,shirai20}. In the presence of additional symmetries, e.g., for a weakly open integrable evolution, $\rho_{\boldsymbol\lambda}(t)$ takes the form of a time dependent generalized Gibbs ensemble (GGE) pametrized with $N_C>1$  Lagrange parameters \cite{lange17,lange18,reiter21,schmitt22,ulcakar23,ulcakar24,bouchoule20,rossini21,gerbino23,perfetto23,rowlands23,riggio24,starchl22,starchl24, lumia24}.
Due to the weak coupling to baths, energy and other conserved quantities are only approximately conserved and are slowly changing due to the dissipation, $\ave{\dot{C}_i} = \tr[C_i \cl{D} \rho(t)] \approx \tr[C_i \cl{D} \rho_{\boldsymbol\lambda}(t)]$. Equivalently, the dissipator super-operator governs the time evolution of the temperature and other Lagrange parameters~\cite{lange18},
\begin{equation}\label{eq:dlambda}
\dot\lambda_i(t) = - \sum_{i'=1}^{N_C}\chi_{ii'}^{-1}(t) \, \tr[C_{i'} \cl{D}\rho_{\boldsymbol\lambda}(t)],
\end{equation}
where $\chi_{ii'}(t) = \tr[C_i C_{i'}\rho_{\boldsymbol\lambda}(t)]-\tr[C_i\rho_{\boldsymbol\lambda}(t)]\tr[C_{i'}\rho_{\boldsymbol\lambda}(t)]$. \new{Since the first order time derivatives of Lagrange parameters are proportional to $\epsilon$, they change on timescale $\epsilon^{-1}$.}

{\it Methods.}
Without assuming any perturbative expansion in small $\epsilon$,  time evolution of  density matrices $\rho(t)$  are obtained by performing a tensor-network (TN) simulation of the Lindblad equation Eq.~\eqref{eq::liouvillian} with open boundary conditions. We employ 4th order time-evolving block decimation for the vectorized density matrix \cite{zwolak04} on system sizes $80 \le L \le 160$, using bond dimension $160 \le \chi \le 240$ and time step $0.05 \le dt \le 0.2$.
In the weak-coupling limit, the G(GE) approximation to the density matrix Eq.~\eqref{eq:rho0} is obtained by evaluating Eq.~\eqref{eq:dlambda}. For the chaotic Hamiltonian without other conserved quantities, Eq.~\eqref{eq:dlambda} reduces to an equation of motion for the inverse temperature $\dot\beta(t) = f(\beta(t))$. We compute this on $L=80$ sites using a tensor-network representation, with thermal states $\rho_{\beta}(t)$ obtained via imaginary-time evolution \cite{schollwock11}. 
For the integrable case, we consider a non-interacting Hamiltonian $H$, for which Eq.~\eqref{eq:dlambda} can be evaluated in thermodynamically large systems by working in the free-fermion representation.
As noted in Ref.~\cite{ulcakar24} and the Supplemental Material (SM) \cite{sm}, $\rho_{\boldsymbol\lambda}(t)$ is then parametrized by time-dependent chemical potentials, which act as Lagrange parameters for the nearly conserved quasiparticle occupations. Their time dependence is calculated from the scattering equation, derived in Ref.~\cite{ulcakar24} and in SM \cite{sm}, capturing  scattering/creation/annihilation of quasiparticles due to the Lindblad perturbation. In the fermionic formulation we employ periodic boundary conditions. However, since our Lindblad operators do not induce currents, the choice of boundary conditions is not essential for sufficiently large system sizes.

We detect the quantum Mpemba effect by comparing distances between time-evolved states $\rho(t)$ and the steady state $\rho_{\infty}$. This can be done either for the full density matrix or the zeroth order (G)GE obtained as described above. The Mpemba effect manifests itself when a state initialized farther from the steady state approaches it faster than one initialized closer.
There is some debate regarding which definition of distance should be used \cite{ares2025}. While the trace distance, $\tr[\sqrt{(\rho(t)-\rho_{\infty})^2}]$, might formally be the most appropriate, it is not suitable for calculations with tensor networks. We employ here a normalized Frobenius norm \cite{fagotti13gge2}, evaluated on reduced density matrices of a subsystem $A$ (chosen in the middle of the system) of compact support $\ell=|A|$, 
\begin{equation}\label{eq:d_norm}
d_{\ell}(\rho(t),\rho_{\infty}) 
= \sqrt{\frac{\tr[(\rho_{A}(t)-\rho_{\infty,A})^2]}{\tr[(\rho_{A}(t))^2] + \tr[(\rho_{\infty,A})^2]}},
\end{equation}
where we introduce the notation $\rho_{A}(t) = \tr_{\bar A}[\rho(t)]$ and
$\rho_{\infty,A} = \tr_{\bar A}[\rho_{\infty}]$. This norm can be readily implemented within the matrix product state representation of $\rho(t)$ and thermal states $\rho_{\beta}(t)$. It is likewise straightforward to compute in the free-fermion Gaussian representation of the GGE $\rho_{\boldsymbol\lambda}(t)$, where the correlation matrix is employed \cite{Fagotti_2010}; see also SM \cite{sm}. In the context of the Mpemba effect, Ref.~\cite{bhore2025} previously compared the behavior of the normalized Frobenius distance \eqref{eq:d_norm} with that of the trace distance. Our comparison for small system sizes, shown in End Matter, similarly supports the use of the normalized norm \eqref{eq:d_norm} over the bare Frobenius distance.

{\it Results.} 
We first test the extreme cases by considering the transverse field Ising Hamiltonian
\begin{equation}\label{eq:H}
H = \sum_j J \sigma^x_j \sigma ^x_{j+1} + h_z \sigma ^z_j + h_x \sigma ^x_j,    
\end{equation}
for (a) chaotic parameters $J=0.75$, $h_z=1.0$, $h_x=0.3$, when there are no additional conserved quantities besides the Hamiltonian itself ($N_C=1$), and (b) integrable parameters $J=0.75$, $h_z=1.0$, $h_x=0$, when the Hamiltonian commutes with macroscopically many conserved quantities ($N_C\sim 2L$).
We use the dissipator \eqref{eq::liouvillian} with
\new{$L_j = \sigma_j^{+}\sigma_{j+1}^{-} + \frac{1}{2}(\sigma_j^{z} + \mathbb{1}_j)$, where $\sigma_j^{\pm} = (\sigma_j^{x} \pm i \sigma_j^{y})/2$}. 

\begin{figure}    
\includegraphics[width=\columnwidth]{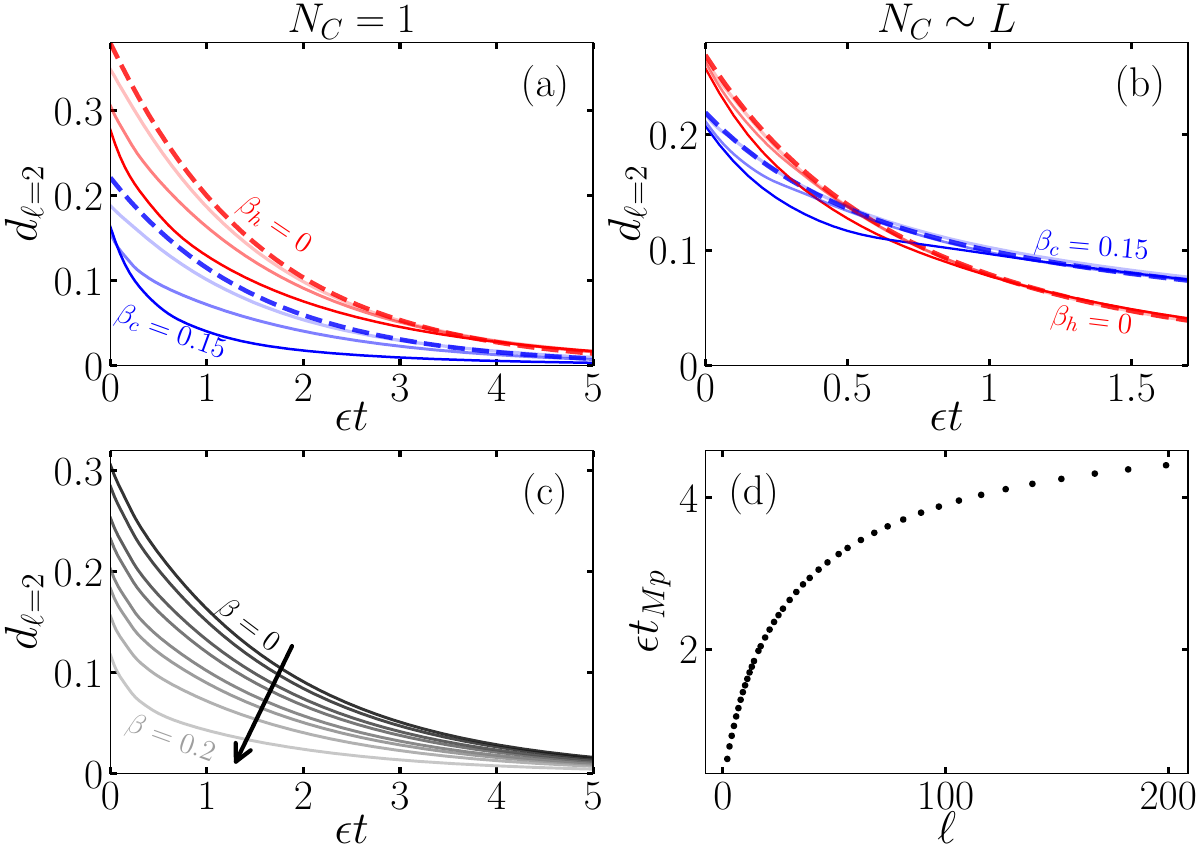}
\caption{(a,b) Distances $d_{\ell=2}(\rho(t),\rho_\infty)$ as a function of rescaled time $\epsilon t$ for the initial thermal states with $\beta_c=0.15$ (blue) and $\beta_h=0$ (red) inverse temperature  for (a)~chaotic and (b)~integrable transverse field Ising Hamiltonian $H$. Results for $\epsilon=0.05,0.2,0.5$ (shown in solid lines of diminishing shades for diminishing $\epsilon$) converge to the (a) Gibbs and (b) GGE results (shown with dashed lines).
(c)~Distances $d_{\ell=2}(\rho(t),\rho_\infty)$ for different initial temperatures $\beta\in[0, 0.2]$ and chaotic $H$ at $\epsilon=0.2$: no Mpemba effect is observed.
\new{Parameters: $L=80$ (chaotic TN results) and $L=160$ (integrable TN results), with $160 \le \chi \le 240$.}
(d)~Crossing times $t_{\text{Mp}}$ as a function of subsystem size $\ell$ for the integrable $H$ extracted from $\rho_{\boldsymbol\lambda}(t)$ calculated on $L=400$.}
\label{figure:fig2}
\end{figure}

Fig.~\ref{figure:fig2}(a,b) shows distance $d_{\ell}(\rho(\epsilon  t),\rho_\infty)$, Eq.~\eqref{eq:d_norm}, as a function of the rescaled time $\epsilon t$ for the initial thermal states with a colder $\beta_c=0.15$ (blue) and a hotter $\beta_h=0$ (red) inverse temperature on support $\ell=2$ for (a)~chaotic $H$ and (b)~integrable $H$. According to our distance measure \eqref{eq:d_norm}, the hotter initial state is farther away from the steady state. Solid lines denote different coupling strengths, $\epsilon=0.05, 0.2, 0.5$, with lighter shades corresponding to smaller $\epsilon$, while dashed lines are extracted from the Gibbs and GGE approximation of the time evolution. In the rescaled time $\epsilon t$, the finite coupling results indeed approach the (G)GE prediction $d_{\ell}(\rho_{\boldsymbol{\lambda}}(\epsilon t),\rho_{\boldsymbol{\lambda},\infty})$, with the distance \eqref{eq:d_norm} applied to the density matrices calculated from Eqs.~(\ref{eq:rho0},\ref{eq:dlambda}), confirming that Gibbs and generalized Gibbs ensembles are the appropriate zeroth-order constituents of the expansion of weakly dissipative dynamics. 
Panels (a) and (b) suggest the absence of the Mpemba effect in the chaotic  $N_C=1$ case and its presence for the integrable $N_C\sim L$ case. 
In panel (c), we scan through different inverse temperatures $\beta\in[0,0.2]$ of the initial state at $\epsilon=0.2$ for the chaotic Hamiltonian, confirming the absence of a Mpemba effect when no additional conserved quantities are present. In panel (d), we show the dependence of the crossing time $t_{\text{Mp}}$ in distances $d_{\ell}(\rho_{\boldsymbol{\lambda}}(\epsilon t),\rho_{\boldsymbol{\lambda},\infty})$ for the integrable $H$ on support $\ell$ calculated with the GGE approximation on $L=400$. 
For our choice of parameters the crossing times remain $\mathcal{O}(1)$ at all $\ell$. 
\new{At least for cases with the Mpemba effect occurring close to the steady state and on large enough $L$, the crossing time saturates with $\ell$, see SM \cite{sm}. 
In general, experimentally obtaining distances computed from tomographically reconstructed reduced density matrices is more feasible for small values of $\ell$.}

From all cases considered, a consistent picture emerges: for the integrable, weakly open dynamics, the Mpemba effect is observed by measuring  distances between the time-propagated state and the steady state, regardless of whether the system is coupled to the baths with an infinitesimal or finite strength. We discuss in the End Matter the interval of initial temperatures within which the Mpemba effect is observed. On the other hand, the Mpemba effect is generically not observed for the chaotic weakly open dynamics with a single approximately conserved quantity.

As underlined in the equation of motion \eqref{eq:dlambda}, within the Gibbs ensemble approximation for chaotic Hamiltonians, the dissipative dynamics are constrained to a one-dimensional manifold of thermal states. 
If the dynamics starting from two different thermal states were to intersect at a Gibbs ensemble with the same temperature, both would then follow the same temperature trajectory, $\beta(t)$. This, however, does not occur: the evolution from different initial states is smooth and monotonic and the trajectories do not cross while relaxing toward the steady-state temperature (Fig.~\ref{figure:fig2}c). \new{To avoid false detection of the Mpemba effect for thermal states, it is important to chose a distance which is a monotonic function of the absolute temperature distance to the steady-state temperature $|\beta - \beta_{\infty}|$ \cite{lr-17}. We check that this is true for the distance \eqref{eq:d_norm} and chaotic Ising Hamiltonian.
In contrast to the chaotic case,} for an integrable Hamiltonian, Eq.~\eqref{eq:dlambda} can be interpreted as an equation of motion in a multidimensional space of Lagrange parameters, spanning the manifold of all possible generalized Gibbs ensembles. Even when starting from a thermal one-dimensional sub-manifold, weakly dissipative evolution under the integrable Hamiltonian will generically take the dynamics into the higher-dimensional GGE manifold, Fig.~\ref{figure:fig1}. 
Calculating the distance between the time-dependent GGE and the steady state projects the evolution from a multi-dimensional space to a single dimension, possibly allowing for crossing between trajectories corresponding to different initial states. The above arguments are based on equations of motion for the infinitesimally small coupling to baths. Importantly, our tensor network calculations at finite coupling $\epsilon$ support similar behavior also when the exact dynamics contains some small deviations from the thermal and generalized Gibbs ensembles.
\new{In the End Matter, we give further evidence for the stability of the Mpemba effect at finite $\epsilon$ and integrable Hamiltonian.}

\begin{figure}    
\includegraphics[width=\columnwidth]{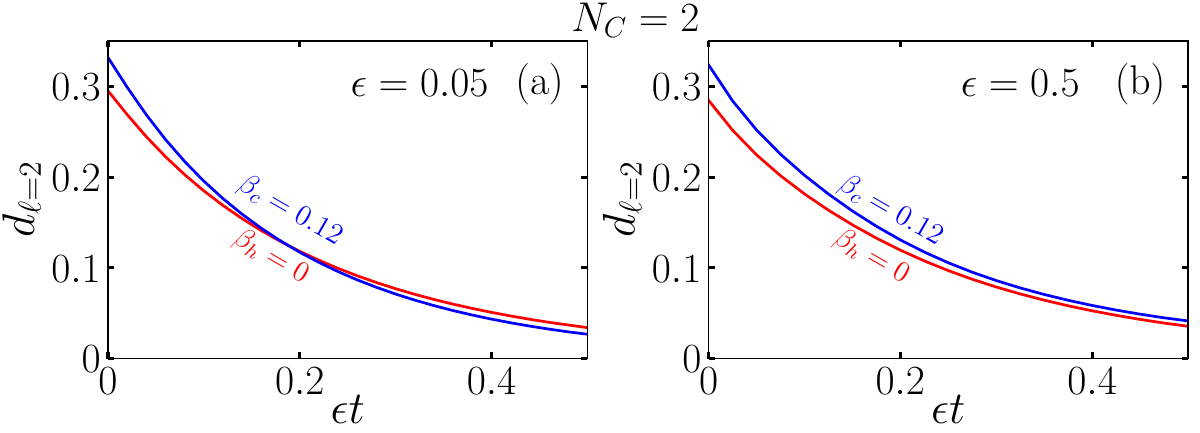}
\caption{Distances $d_{\ell}(\rho(t),\rho_\infty)$ as a function of rescaled time $\epsilon t$ for the initial thermal states with $\beta_c=0.12$ (blue) and $\beta_h=0$ (red) inverse temperature and chemical potential $\mu(0)=0$ for chaotic Hamiltonian with two conserved quantities, Eq.~\eqref{eq:H2}, at (a) $\epsilon=0.05$ and (b) $\epsilon=0.5$. Parameters: $L=80$, $\ell=2$, $\chi=200 \text{ (a)}, 160 \text{ (b)}$.}
\label{figure:fig3}
\end{figure}
While having one or macroscopically many conserved quantities are the extreme cases, one might ask whether having just two approximately conserved quantities is sufficient to observe the Mpemba effect in weakly dissipative dynamics. To examine this case, we consider 
\begin{equation}\label{eq:H2}
H = - \sum_j J (\sigma^x_{j} \sigma^x_{j+1} + \sigma^y_{j} \sigma^y_{j+1}) + \Delta_j\sigma^z_{j} \sigma^z_{j+1}
\end{equation}
with $J=1$, $\Delta_j = 1.2  + 0.4 \cdot(-1)^j$. The Hamiltonian \eqref{eq:H2} is not integrable, but has as an additional conserved quantity, the total magnetization $S^z = \sum_j \sigma^z_j$. 
As a dissipative perturbation, we consider Lindblad operators
\new{$L_i=\frac{1}{2}\sigma_i^+ (\mathbb{1}_{i+1}-\sigma_{i+1}^z) +  \sigma^x_i$}, again coupled in the bulk. In this case, the steady state and the dissipative dynamics can be approximately described with a time dependent temperature and a chemical potential $\mu(t)$, associated with the approximately conserved magnetization. Importantly, we need a Lindblad perturbation that mixes magnetization sectors, otherwise we have a separate steady state in each sector and dynamics within a sector is as if we had a single approximately conserved quantity. 
Fig.~\ref{figure:fig3}(a) shows that the Mpemba effect is less pronounced, yet still visible, e.g., for initial temperatures $\beta_c=0.12$, $\beta_h=0$, chemical potential $\mu(0)=0$ and the above choice for the dissipator. We note that a stronger dependence on the coupling strength to the baths is observed, with the Mpemba effect occurring only for sufficiently small $\epsilon \lesssim 0.2$; see Fig.~\ref{figure:fig3}(b) lacking the Mpemba effect at $\epsilon=0.5$. 
\new{In the End Matter, we further investigate the observed $\epsilon$ dependence compared to the integrable case.}
Also, a finer tuning of the Lindblad operators appears to be required for the effect to manifest. 

Unlike previous examples, we see from Fig.~\ref{figure:fig3} that the initial state of the hotter infinite temperature $\beta_h=0$ is actually closer to the steady state than the colder $\beta_c=0.12$. We are thus actually observing the so called `inverse Mpemba effect' \cite{lr-17,kcb-22,shapira}. However, for our model this is not such a surprise: while the Hamiltonian \eqref{eq:H2} has ferromagnetic interactions, the Lindblad operator prefers antiferromagnetic order, present in negative temperature states.

\begin{figure}    
\includegraphics[width=\columnwidth]{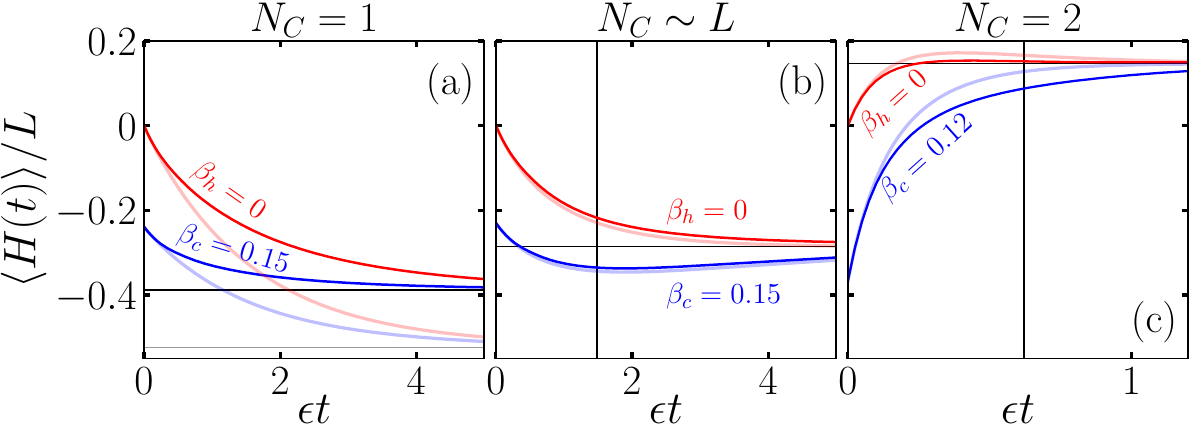}
\caption{Energy density $\ave{H}/L$ as a function of rescaled time $\epsilon t$ for (a)~the integrable Hamiltonian \eqref{eq:H} with $N_C\sim L$ and (b)~the chaotic Hamiltonian with a single additional conserved quantity \eqref{eq:H2}, $N_C=2$. Parameters:  $\epsilon=0.05,0.2,0.5$, $L=160$, $180\le \chi\le 240$ ($N_C \sim L$) and $\epsilon=0.05,0.2,0.5$ $L=80$, $160\le \chi\le 200$ ($N_C=2$).}
\label{figure:fig4}
\end{figure}

{\it Discussion.}
Analysis on the level of distances is theoretically appropriate, but experimentally \new{more demanding}. In Fig.~\ref{figure:fig4}, we show how the quantum Mpemba effect \new{can} manifest itself in terms of observables. We show \new{time dependence of} energy density for \new{the three Hamiltonians considered above with (a) $N_C=1$, (b) $N_C\sim L$, and (c) $N_C=2$ conserved quantities for $\epsilon=0.05$ (fainter) and $\epsilon=0.5$ (darker) coupling strenghts.
In panels (b) and (c), the Mpemba effect does not manifest itself in a crossing. Rather, the energy density that was initially further from the steady state value (denoted with the horizontal line) gets at some finite time closer to it (for $\epsilon=0.05$ denoted with the vertical line). In panel (a) such an effect is absent.}
\new{For considered cases at thermal initial states, energy thus correlates with the behaviour of the distance \eqref{eq:d_norm}, however, as discussed in SM \cite{sm}, this is not generically true. If the Mpemba effect happens at asymptotically long times, distance \eqref{eq:d_norm} can be reconstructed  by combining experimentally measured conserved quantities and classical calculations, see SM \cite{sm}}.

The occurrence of the Mpemba effect in dissipative systems is often related to the orthogonality of the initial state to the slowest decaying mode \cite{cll-21,goold,kcl-22,bh-22,goold2,Hayakawa0,Hayakawa1,shapira,ias-23, zhou-23, liu-24,fs-24,mcblg-24, gsm-24,westhoff2025}.
An initial state that has zero overlap with the slowest decaying mode will decay at long times faster compared to states that have a finite overlap. The so called \textit{strong} Mpemba effect can occur when such an initial state is at the same time farther away from the steady state than another non-orthogonal initial state.
In the End Matter, we corroborate the main-text-results on large system sizes with the exact diagonalization study of orthogonality to the slowest mode. This analysis helps us identify the interval of initial temperatures exhibiting the Mpemba effect and pinpoint the condition for the strong effect to occur.
\new{When the Mpemba effect occurs at asymptotically long times, one can linearize the dynamics around the steady state and formulate a simple geometric condition for its occurrence, reported in SM \cite{sm}.}

{\it Conclusions.}
In this paper, we studied the Mpemba effect in weakly open quantum many-body systems initialized in different thermal states. We give numerical evidence that the Mpemba effect can occur only when the underlying Hamiltonian has additional (local) conserved quantities. If energy is the only approximately conserved quantity, the dynamics can be approximated as an evolution between Gibbs ensembles of different temperatures. In the presence of additional conserved quantities and for integrable Hamiltonians, additional Lagrange parameters lead to the state being approximated with time dependent generalized Gibbs ensembles. When projecting this dynamics on one-dimensional distances between the time-dependent state and the steady state, crossings can appear if the state evolves in the multidimensional GGE manifold and cannot appear for dynamics within the one-dimensional thermal manifold. 

Even though our argument applies at the level of the zeroth-order approximation, 
tensor-network simulations of the full density matrix \new{and Liouville spectrum analysis (End Matter)} suggest our claim is valid for finite, experimentally relevant bath couplings.
\new{The appearance of the Mpemba effect is consistent with previous studies in dissipative systems where the overlap with the slowest decaying mode plays a crucial role (End Matter). Using this, we identify possible initial (thermal) states exhibiting the effect and give a compact geometric condition for it to occur at late time (SM \cite{sm}).}

\new{
While the presence of additional approximate conserved quantities appears to be a necessary condition for observing the Mpemba effect, it is not sufficient, and a general criterion for the choice of Lindblad operators remains to be determined. Furthermore, the stability of the Mpemba effect at finite system-bath coupling, as well as its dependence on the number of conserved quantities, requires further investigation. In this context, one promising direction would be to examine whether (approximate) integrability can generate sufficiently nonthermal transient dynamics even in the presence of thermal baths.} 
Another promising direction is to relate our arguments to the occurrence of the Mpemba effect in classical systems, in connection with their symmetries. Finally, our general conclusions may facilitate experimental verification using  cold-atom quantum simulators, for example, considering loss of atoms. 

\begin{acknowledgments}
We thank S. Murciano, P. Calabrese and G. Teza for useful discussions. I.U, G.L. and Z.L acknowledge the support by P1-0044 program of the Slovenian Research and Innovation Agency (ARIS), European Union Horizon 2020 under the QuantERA II project QuSiED (No 101017733), and ERC StG 2022 project DrumS by Horizon Europe, Grant Agreement 101077265. R.S.
acknowledges support from ERC Advanced grant No. 101096208– QUEST, and Research Programme P1-0402
of Slovenian Research and Innovation Agency (ARIS). Calculations were performed at the cluster `spinon' of JSI, Ljubljana, using QuSpin \cite{QuSpin1} and ITensors.jl \cite{itensor}. 
\end{acknowledgments}

\bibliography{biblio,iterative_iris}

\section{End Matter}
The Quantum Mpemba effect in dissipative systems is in most cases related with the orthogonality to the slowest decaying mode of the Liouvillian \cite{cll-21,goold,kcl-22,bh-22,goold2,Hayakawa0,Hayakawa1,shapira,ias-23, zhou-23, liu-24,fs-24,mcblg-24, gsm-24,westhoff2025}. Namely, initializing the system in a state that is orthogonal to the slowest decaying mode should result in a faster relaxation to the steady state compared to an initial state with a finite overlap with the slowest mode, which is initially closer to the steady state. 

The explicit construction of the orthogonalization procedure by diagonalizing the Liouvillian is practically difficult and limits the accessible system sizes.
In the main text, we showed that the density matrix dynamics can be approximated with (G)GEs. However, without any further assumptions, following bare perturbative arguments for weak dissipation \cite{lenarcic18}, the slow dynamics happens within the diagonal subspace, spanned by exponentially many projectors on the Hamiltonian eigenstates $\ket{m}\bra{m}$. 
I.e., the slowest mode primarily lies in this manifold and also the initial thermal states are fully embedded in the diagonal subspace. The slow dynamics is captured by the Liouvillian projected onto this subspace of matrices. To perform the orthogonality analysis we thus resort to the diagonalization of the projected Liouvillian, represented by the matrix $\mathfrak{D}_{mn} = \ave{m|(\cl{D}|n\rangle \langle n|)|m}$. This allows us to study somewhat larger system sizes $L\le 16$, compared to the analysis that would be performed on the non-projected Liouvillian.

\begin{figure}    
\includegraphics[width=\columnwidth]{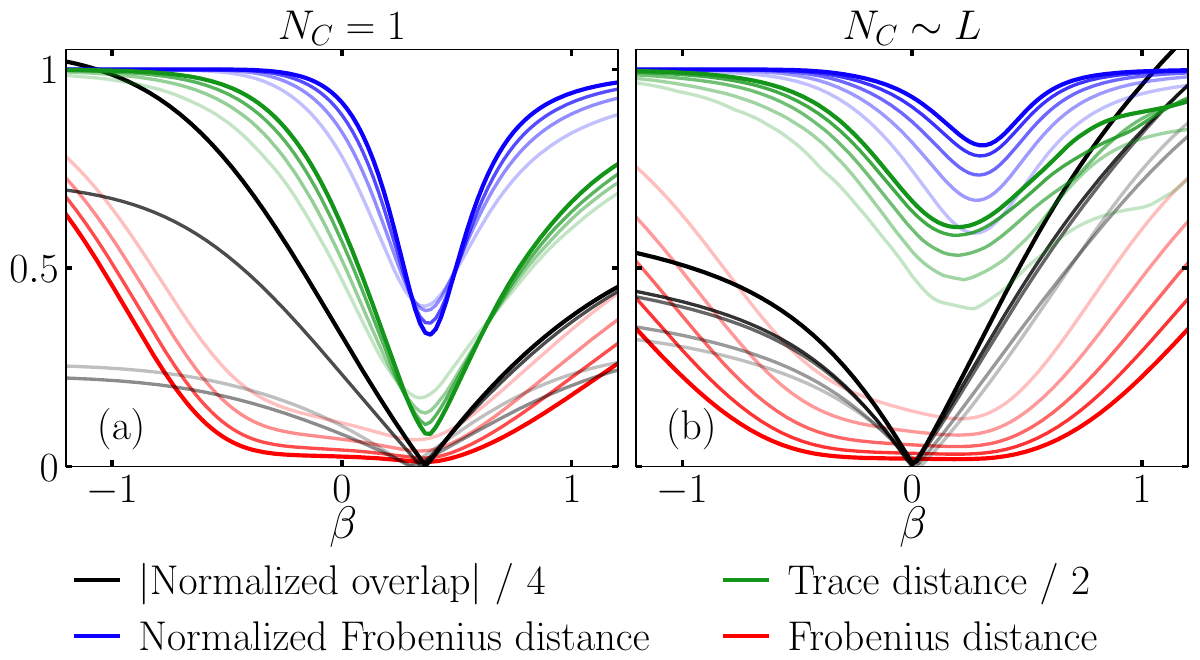}
\caption{ 
    Normalized overlap of different thermal states with the slowest mode and different distances (trace distance, Frobenius distance, normalized Frobenius distance) of thermal states to the steady state for (a) chaotic and (b) integrable transverse field Ising parameters used in the main text. System sizes (a) $L=8, 10, 12, 14$ and (b) $L=10, 12, 14, 16$ from lighter to darker colors.
    }
    \label{figure:figEM1}
\end{figure}

Fig.~\ref{figure:figEM1} shows normalized overlaps of different initial thermal states $\rho_\beta$, characterized by the inverse temperature $\beta$, and the left eigenvector $\rho_{\text{slow}}^{(l)}$, corresponding to the slowest right eigenvector $\rho_{\text{slow}}^{(r)}$ of the projected Liouvillian, $||\rho_{\text{slow}}^{(r)}||_1*\tr[(\rho_{\text{slow}}^{(l)})^\dagger \rho_\beta]$. Normalization is chosen to avoid any ambiguity regarding the separate normalization of the left and right eigenvectors, under the standard assumption $\tr[(\rho_{i}^{(l)})^\dagger  \rho_{j}^{(r)}]=\delta_{ij}$. Fig.~\ref{figure:figEM1}(a) shows result for the chaotic choice of transverse field Ising parameters, while panel (b) shows results for the integrable parameters. For both cases, there exists a thermal initial state that is orthogonal to the slowest mode. Different shades correspond to different system sizes, from lighter to darker shade for increasing system sizes $L=8, 10, 12, 14, 16$. Whether this can allow for the occurrence of the Mpemba effect depends on the distance of the initial thermal state to the steady state, on Fig.~\ref{figure:figEM1} shown for different definitions of distance: the trace distance $d_T(\rho_\beta,\rho_{\infty}^{(D)}) = \tr\Big[\sqrt{(\rho_\beta-\rho_{\infty}^{(D)})^2}\Big]$, the Frobenius distance $d_F(\rho_\beta,\rho_{\infty}^{(D)})=\sqrt{\tr[(\rho_\beta-\rho_{\infty}^{(D)})^2]}$
and the normalized Frobenius distance 
$d_L(\rho_\beta,\rho_{\infty}^{(D)})=\sqrt{\tr[(\rho_\beta-\rho_{\infty}^{(D)})^2]/(\tr[(\rho_\beta)^2] + \tr[(\rho_{\infty}^{(D)})^2])}$. Here we consider the steady state projected onto the diagonal subspace, $\rho_{\infty}^{(D)}$. Again, increasingly darker shades correspond to increasingly larger system sizes. The most important conclusion from the above analysis is that for the chaotic model with a single conserved quantity, the initial thermal state that is orthogonal to the slowest mode is also the state that is closest to the steady state, ruling out the observation of the Mpemba effect. With increasing system sizes minima in different distances are just deepening, since in the thermodynamic limit the diagonal projection of the steady state is also a thermal state and thus orthogonal to the other decaying modes. On the other hand, for the integrable model in panel (b), the dip in the distances is at a different temperature and is getting shalower with increasing system sizes where the GGE character of the steady state is becoming more pronounced. In the interval of initial state temperatures between the minima of the overlap with the slowest mode and the distance to the steady state are pairs of initial temperatures for which the Mpemba effect could be observed. 
If one of them is the thermal state orthogonal to the slowest mode (in our case at or close to $\beta = 0$), we observe the \textit{strong} Mpemba effect.
A secondary observation is that various distances behave qualitatively similar but quantitatively different. Trace and normalized Frobenius distance are somewhat more similar, which is the reason why we used the latter also in the main text. Previously it was already pointed out by Ref.~\cite{fagotti13gge2} that the normalized Frobenius distance captures the comparison of observables and reduced density matrices in the two states considered better than the bare Frobenius distance.
Other studies have used the entanglement asymmetry \cite{alba2017, joshi-24, murciano2024}, the trace distance \cite{shapira, zhang2024}, the Frobenius distance \cite{cll-21}, relative entropies \cite{goold} and reduced fidelities \cite{parez} to quantify the quantum Mpemba effect.

\begin{figure}    
\includegraphics[width=\columnwidth]
    {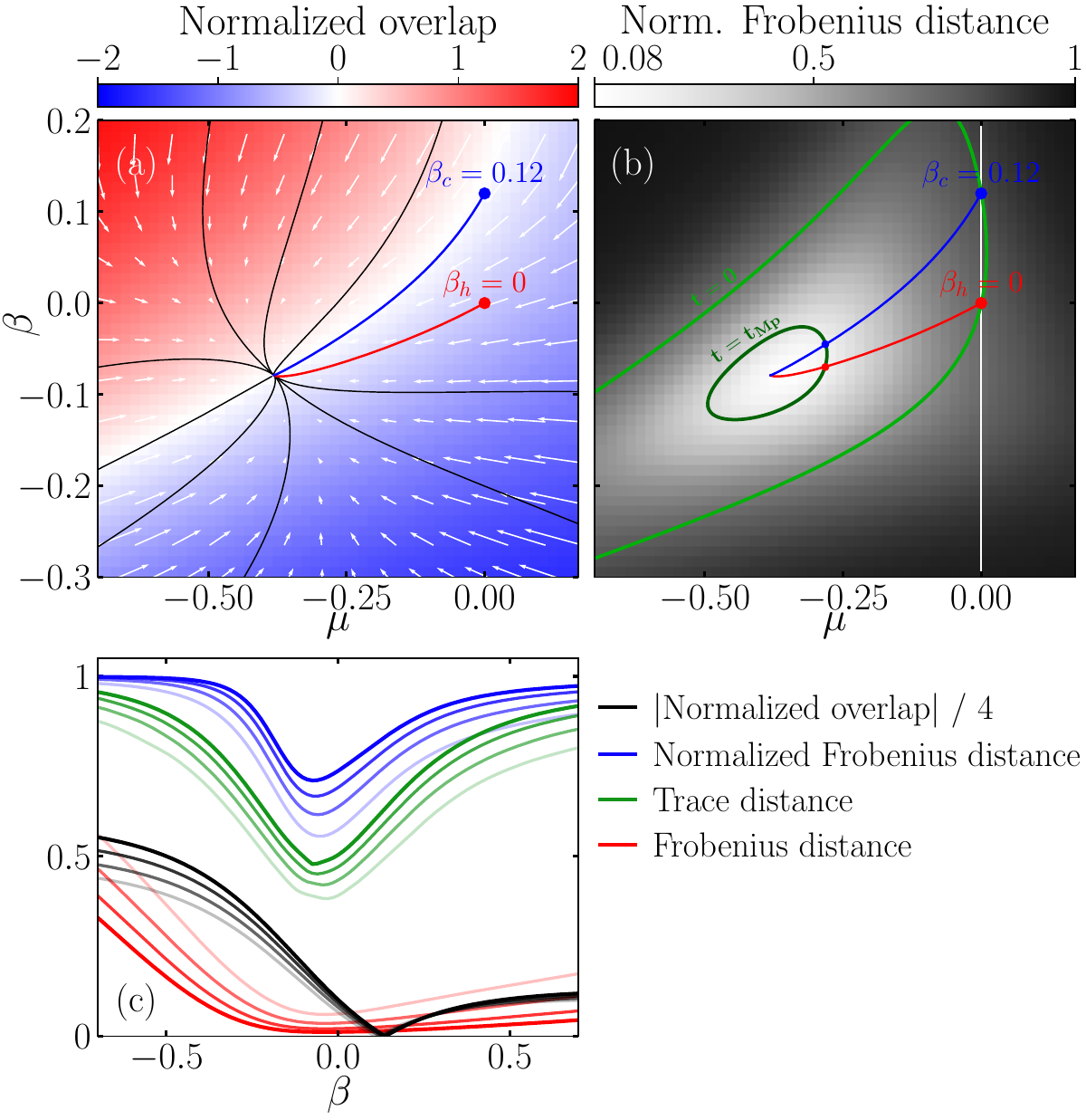}
    \caption{ 
    Orthogonality analysis for model \eqref{eq:H2}. (a)~Normalized overlap of states parametrized with inverse temperature $\beta$ and chemical potential $\mu$ with the slowest mode. \new{White arrows denote time derivatives $(\dot{\beta}, \dot{\mu})$ at different points. Full lines represent several trajectories $(\beta(\epsilon t), \mu(\epsilon t))$, and the blue and red line start at thermal states.} (b)~Trace distance between the steady state and the initial states parametrized with inverse temperature $\beta$  and chemical potential $\mu$. \new{Light green curve denotes all initial states, which will have the same normalized Frobenius distance to the steady state $d_L = 0.233$ at time $t_{Mp} = 0.32$ (denoted by dark green).} System size $L=14$.  (c) Normalized overlap with the slowest mode and different distances (trace distance, normalized Frobenius distance and Frobenius distance) of initial thermal states ($\mu=0$) to the steady state, $L=8, 10, 12, 14$ from lighter to darker colors.}
    \label{figure:figEM2}
\end{figure}

Next, we perform a similar analysis for the chaotic model \eqref{eq:H2} with Lindblad operators \new{$L_i=\frac{1}{2}\sigma_i^+ (\mathbb{1}_{i+1}-\sigma_{i+1}^z) +  \sigma^x_i$}, shown in the main text as an example of a setup with two approximately conserved quantities. Here, we perform the same analysis as in Fig.~\ref{figure:figEM1}, but for the initial states $\rho_{\beta,\mu}$ characterized with inverse temperature $\beta$ and chemical potential $\mu$, associated with total $S^z$. Panel~(a) shows contour plots \new{(shades of blue and red)} for the normalized overlap $||\rho_{\text{slow}}^{(r)}||_1*\tr[(\rho_{\text{slow}}^{(l)})^\dagger \rho_{\beta,\mu}]$ with the slowest mode, while panel~(b) shows \new{(in shades of gray) normalized Frobenius} distance of such initial states to the diagonal projection of the steady state \new{$d_L(\rho_{\beta,\mu},\rho_{\infty}^{(D)})$} for $L=8, 10, 12, 14$. Again, there are pairs of parameters $(\beta,\mu)$ for which the initial state is (nearly) orthogonal to the slowest mode and yet further away from the steady state. Pairs of initial states shown in the main text in Fig.~\ref{figure:fig3} are denoted with dots of the corresponding color. Panel~(c) focuses on the initial states with $\mu=0$ and shows the normalized overlap of inital $\rho_{\beta,\mu=0}$ with the slowest mode, the trace $d_T(\rho_{\beta,\mu=0},\rho_{\infty}^{(D)})$, the Frobenius distance $d_F(\rho_{\beta,\mu=0},\rho_{\infty}^{(D)})$ and the normalized Frobenius distance $d_L(\rho_{\beta,\mu=0},\rho_{\infty}^{(D)})$ as a function of $\beta$. Orthogonality condition around $\beta=0.12$ motivates our choice presented in the main text, $\beta_c=0.12$ and $\beta_h=0$.

\new{In Fig.~\ref{figure:figEM2}(a) we additionally plot the flow field defined as time derivatives of the Lagrange parameters $(\dot{\mu},\dot{\beta})$, Eq.~\eqref{eq:dlambda}, at different points in the GGE manifold. By integrating these equations of motion from some initial point in the manifold $( \mu_0, \beta_0)$, we obtain a \textit{trajectory} $(\mu(\epsilon t), \beta(\epsilon t))$. Different trajectories are denoted with full lines. Trajectories for $(0,\beta_c)$ and $(0,\beta_h)$ considered in the main text 
are denoted with a blue and red line. All trajectories converge to the steady state $(\mu_{\infty}, \beta_{\infty})$.}

\new{
In the main text we focused on thermal initial states, with a pair of them possibly showing a crossing at a certain time $\epsilon t_{Mp}$. The trajectory interpretation helps us recognize that for more general initial states with all $N_C$ Lagrange parameters finite, there is a $N_C-1$ dimensional (deformed) ellipsoid of initial states that all cross at the same distance and time. For $N_C = 2$, the dark green anisotropically deformed ellipse in Fig.~\ref{figure:figEM2}(b) marks the set of states at distance $d_L(\rho_{\beta,\mu}, \rho_{\infty}^{(D)}) = 0.233$ from the steady state, at which the two thermal states $\beta_c=0.12$ and $\beta_h=0$ from the main text (red and blue point) intersect. By backpropagating the trajectories, we reconstruct the set of all states (shown in bright green) that intersect at the same distance at time $\epsilon t_{Mp} = 0.32$.
}

\begin{figure}
\includegraphics[width=\columnwidth]
    {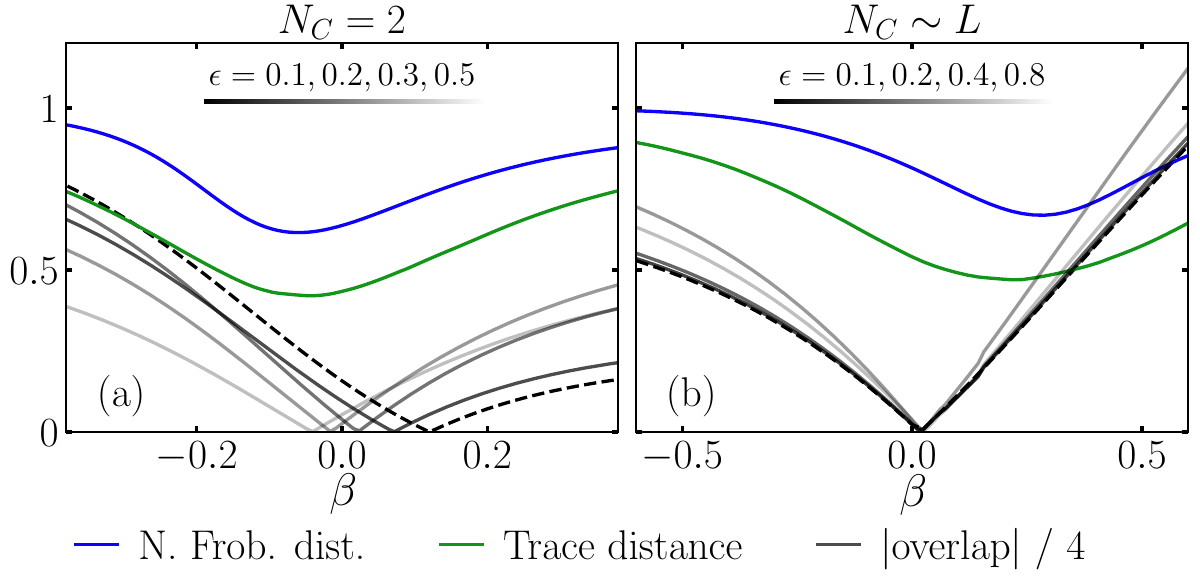}
    \caption{ 
    \new{Orthogonality analysis for (a) the model with two conserved quantities \eqref{eq:H2} and (b) the integrable model \eqref{eq:H} at finite values of coupling $\epsilon$. The normalized overlap is plotted, with the black lines becoming increasingly transparent as $\epsilon$ increases. Both distances and the dashed black line show the results of the diagonal ensemble analysis used in Figs. \ref{figure:figEM1} and \ref{figure:figEM2}. System size $L = 10$.}}
    \label{figure:figEM3}
\end{figure}

\new{Finally, we look at the dependence and the breakdown of the Mpemba effect at finite coupling strength to the baths $\epsilon$. We perform a similar analysis of the normalized overlap and the distances to the steady state for different thermal initial states, as done in Figs. \ref{figure:figEM1} and \ref{figure:figEM2}, except that we now use the exact Liouvillian \eqref{eq::liouvillian} and find the leading eigenvectors using the Arnoldi iteration. 
We consider the slowest mode having a non-vanishing overlap with initial thermal states and plot it with gray lines in Fig. \ref{figure:figEM3} for a few $\epsilon$ (increasing $\epsilon$ is denoted with increasingly transparent lines, the diagonal ensemble results $\epsilon\to 0$ is denoted with a dashed line). 
The distances of thermal states to the steady state are plotted only for the diagonal ensemble as their change with finite $\epsilon$ is negligible. On the other hand, the change in normalized distance is apparent. As seen in Fig.~\ref{figure:figEM3}(a), in the case of two conserved charges, increasing $\epsilon$ causes the minimum to significantly drift towards the minimum of the distances, thereby reducing the range of initial states exhibiting the Mpemba effect. For $\epsilon = 0.5$ the window has almost completely closed up. On the other hand, in the case of the integrable model, Fig.~\ref{figure:figEM3}(b), the overlap with the slowest relevant mode is very stable towards finite $\epsilon$, making in turn the Mpemba effect stable as well. 
}

\newpage
\newpage
\newpage

\renewcommand{\thetable}{S\arabic{table}}
\renewcommand{\thefigure}{S\arabic{figure}}
\renewcommand{\theequation}{S\arabic{equation}}
\renewcommand{\thepage}{S\arabic{page}}

\renewcommand{\thesection}{S\arabic{section}}

\onecolumngrid

\setcounter{figure}{0}
\setcounter{equation}{0}
\setcounter{page}{0}

\newpage

\begin{center}
{\large \bf Supplemental Material:\\
Conserved quantities enable the quantum Mpemba effect in weakly open systems}\\
\vspace{0.3cm}
Iris Ul\v cakar$^{1,2}$, Rustem Sharipov$^{2}$, Gianluca Lagnese$^{1}$, and Zala Lenar\v ci\v c$^{1}$\\
$^1${\it Department of Theoretical Physics, J. Stefan Institute, SI-1000 Ljubljana, Slovenia} \\
$^2${\it Faculty for Mathematics and Physics, University of Ljubljana, Jadranska ulica 19, 1000 Ljubljana, Slovenia} \\
\end{center}

In the Supplemental Material, we (i) summarize the calculation of dynamics within the generalized Gibbs ensemble approximation in the free fermion language, (ii) explain how to evaluate distances between such Gaussian states, and (iii) linearize equations of motion for Lagrange parameters at asymptotic times and obtain a geometric condition for the Mpemba effect to occur for the normalized Frobenius distance and different observables in this regime.

\vspace{0.6cm}

\twocolumngrid

\label{pagesupp}

\section{Free fermion representation} \label{app:free}
Here, we discuss how to efficiently reformulate the problem for the integrable transverse field Ising model 
\begin{equation}\label{eq:H0spin}
H = \sum_i J \sigma^x_i \sigma ^x_{i+1} + h_z \sigma ^z_i,    
\end{equation}
in terms of fermionic operators, so as to evaluate the equation of motion \eqref{eq:dlambda} in thermodynamically large systems using Wick contractions for Gaussian states. Same treatment has already been performed in Ref.~\cite{ulcakar24}, we repeat it here for completeness.

Non-interacting Hamiltonian \eqref{eq:H0spin} can be diagonalized using a sequence of transformations: the Jordan-Wigner transformation 
$\sigma_j^z = 2c_j^{\dagger}c_j - 1,\
    \sigma_j^{+} = e^{i\pi\sum_{l<j}n_l}c_j^{\dagger},$
the Fourier transform
$c_j = \frac{e^{-i\pi/4}}{\sqrt{L}} \sum_q e^{iqj} c_q.$
and the Bogoliubov transformation $ c_q = u_q d_q - v_qd_{-q}^{\dagger}$
 with
$a_q = 2(J\cos{q}+h_z)$,
$b_q = -2J\sin{q}$,
$u_q = \frac{\varepsilon_q+a_q}{\sqrt{2\varepsilon_q(\varepsilon_q + a_q)}}$,
$v_q = \frac{b_q}{\sqrt{2\varepsilon_q(\varepsilon_q + a_q)}}$.
Finally, the Hamiltonian is written in terms of quasiparticle occupation operators $n_q=d_q^{\dagger}d_q$ and dispersion relation $\varepsilon_q = 2\sqrt{J^2 + 2 h_z J \cos{q} + h_z^2}$, 
as
\begin{equation}\label{eq::disp-2}
    H = \sum_q \varepsilon_q \left(n_q -\frac{1}{2}\right).
\end{equation}

In the fermionic language, the zeroth order generalized Gibbs ensemble approximation to the dynamic can be parametrized with chemical potentials $\mu_q(t)$, associated with occupation operators $n_q$,  
\begin{equation}\label{eq:rho0_f}
    \rho_{\boldsymbol\lambda}(t) = \frac{e^{-\sum_q \mu_q(t) n_q}}{\tr[e^{-\sum_q \mu_q(t) n_q}]}.
\end{equation}
For these calculations, we use periodic boundary conditions in the spin formulation, which are translated to periodic boundary conditions in the fermion picture for an odd number of particles and anti-periodic for an even number of particles. The two parity sectors contain different wave vectors $\mathcal{K}^{+}=\{\frac{2\pi}{L}(q+\frac{1}{2}), q = 0, \dots L-1\}$ (even sector) and $\mathcal{K}^{-}=\{\frac{2\pi}{L}q, q = 0, \dots L-1\}$ (odd sector). 

In order to evaluate the equation of motion (Eq.~\eqref{eq:dlambda} from the main text) for the occupation operator,
\begin{equation}
    \label{eq:terms}
    \ave{\dot{n}_q} = \epsilon \sum_j \tr[ L_j^{\dagger}n_q L_j\rho(t)] -\tr[n_q L_j^{\dagger}L_j \rho(t)]
\end{equation}
we need to express the Lindblad operators considered \new{$ L_j = \sigma_j^{+}\sigma_{j+1}^{-} + \frac{1}{2}(\sigma_j^{z} + \mathbb{1}_j)$, where $\sigma_j^{\pm} = (\sigma_j^{x} \pm i \sigma_j^{y})/2$} in terms of Bogoliubov modes
\begin{equation}\label{eq::Li}
    L_j =  \sum_{q, q'} \frac{e^{-ij(q-q')}}{L}(1+e^{iq'}) (u_qd_q^{\dagger} - v_qd_{-q})(u_{q'}d_{q'} - v_{q'}d_{-q'}^{\dagger}).
\end{equation}
Using Wick contractions, Eq.~\eqref{eq:terms} can be simplified into a compact scattering equation
\begin{align}\label{eq::nqdot}
\langle \dot{n}_q \rangle = \frac{2\epsilon}{L} & \sum_{q'}  
  f^s_{q, q'} \langle 1 - n_q \rangle \langle n_{q'} \rangle 
 - f^s_{q', q} \langle n_q \rangle \langle 1 - n_{q'} \rangle \notag \\
 & + f^c_{q, q'} \langle 1 - n_q \rangle \langle 1 - n_{q'} \rangle - f^a_{q, q'} \langle n_q \rangle \langle n_{q'} \rangle. 
\end{align}
parametrized with positive functions encoding the scattering processes
\begin{align}\label{eq::f1q}
f^s_{q, q'} = &u_q^2 u_{q'}^2 (1+\cos q') + v_q^2 v_{q'}^2(1+\cos q)\\
& - u_q v_q u_{q'} v_{q'} (1+\cos{q'} +\cos q +\cos(q+q') ),\notag
\end{align}
the creation of $q'$ and $q$ modes
\begin{align}\label{eq::fc2q}
f^c_{q', q} = &v_q^2 u_{q'}^2 (1+\cos {q}) + u_q^2 v_{q'}^2(1+\cos q') \\
& - u_q v_q u_{q'} v_{q'} (1+\cos {q'} +\cos q +\cos(q-{q'}) ),\notag
\end{align} 
and annihilation of $q'$ and $q$ modes
\begin{align}\label{eq::fa2q}
f^a_{q', q} = &v_q^2 u_{q'}^2 (1+\cos {q'}) + u_q^2 v_{q'}^2(1+\cos q)\notag  \\
&- u_q v_q u_{q'} v_{q'} (1+\cos {q'} +\cos q +\cos(q-{q'}) ).\notag 
\end{align} 
As written in the main text, Eq.~\eqref{eq:dlambda}, the update of Lagrange multipliers is related to the update of conserved quantities via susceptibility matrix, which is for occupation operators simply diagonal
$\dot{\mu}_q(t) = - \frac{\ave{\dot{n}_{q}}(t)}{\chi_{q,q}(t)}$, $\chi_{q,q}(t)= \frac{e^{-\mu_q(t)}}{(1+e^{-\mu_q(t)})^2}.$

Considered Lindblad dissipators preserve the particle number parity, Eq. \eqref{eq::Li}, therefore the dynamics in the two parity sectors is decoupled at the level of zeroth order approximation $\rho_{\boldsymbol\lambda}(t)$. However, at finite coupling to baths this symmetry is lost. When comparing the GGE results in the main text to finite coupling tensor network ones, we perform averaging over both parity sectors, which are indistinguishable in the thermodynamic limit anyway.

\subsection{Frobenius distance for free fermion representation of density matrices}

We outline the procedure for calculating the distance between reduced density matrices (RDMs) of two Gaussian states, used in the main text on the zeroth order $\rho_{\boldsymbol\lambda}(t)$ approximation of dynamics, Eq.~\eqref{eq:rho0}, for the integrable Hamiltonian. 

Consider a subsystem of interest $A$ consisting of the first $\ell$ spins, located on sites $i = 1, \ldots, \ell$ and the reduced density matrix $\rho_A=\tr_{\bar{A}}[\rho]$ of state $\rho$. It is convenient to represent $\rho_A$ in terms of Majorana fermions:
\begin{equation}
    \rho_A=\frac{1}{2^\ell} \sum_{\mu_i=0,1} \text{Tr} \left[\rho \prod_{i=1}^{2\ell} a^{\mu_i}_{i}\right] \left(\prod_{i=1}^{2\ell} a^{\mu_i}_{i} \right)^{\dagger}
\end{equation}
where $a_{2i}=c_i+c_i^\dagger,~ a_{2i-1}=i(c_i-c_i^\dagger)$ and $c_i$ is a fermionic operator as introduced in the previous section.  
Importantly, the reduced density matrix of a Gaussian state is also Gaussian. A central quantity that characterizes the reduced density matrix is the correlation matrix
 $\Gamma$, defined on subsystem $A$ of length $\ell$ by:
\begin{equation}
    \Gamma_{A,ij}=\text{Tr}[a_i a_j \rho]-\delta_{ij},~~1\leq i,j \leq 2 \ell
\end{equation}
Using Wick’s theorem, the reduced density matrix can be expressed in Gaussian form
\begin{equation}
    \rho_A \sim e^{-\frac{1}{4}a_i W_{A,ij} a_j},~~~\text{where}~~\tanh \left(\frac{W_A}{2}\right)=\Gamma_A 
\end{equation}
Given the correlation matrices, the trace of the product of RDMs can be computed using \cite{Fagotti_2010}:
\begin{equation}\label{eq:det}
\tr[\rho_A  \tilde\rho_A]=\left(\det \left|\frac{1+\Gamma_A \cdot \tilde\Gamma_A}{2}\right|\right)^{\frac{1}{2}}.
\end{equation}
Thus, the Frobenius distance between two reduced density matrices is obtained from substituting Eq. \eqref{eq:det} into the expression
\begin{equation}\label{eq:distance}
    \tr[(\rho_A-\tilde{\rho}_A)^2]
    =\tr[\rho_A^2]+\tr[\tilde\rho_A^2]-2\tr[\rho_A  \tilde\rho_A]
\end{equation}

We now turn from the discussion of distances from two general Gaussian density matrices to those given by the GGE (\ref{eq:rho0_f})  of transverse field Ising model. In that case, the Majorana fermions operators can be linearly expressed in terms of Bogoliubov quasiparticle operators $d_q^{\dagger} d_p$, Eq.~(\ref{eq::disp-2}), with simple expectation value 
\begin{equation}
    \text{Tr} [d_q^{\dagger} d_p \rho_{\boldsymbol\lambda}]=\delta_{q, p} \frac{e^{-\mu_q}}{1+e^{-\mu_q}}. 
\end{equation}
This simplifies the structure of the correlation matrix, which has the block-Toeplitz form
\begin{equation}
\Gamma_A =
\begin{pmatrix}
\Gamma^{(0)} & \Gamma^{(-1)} & \cdots & \Gamma^{(1-\ell)} \\
\Gamma^{(1)} & \Gamma^{(0)} & \cdots & \vdots \\
\vdots & \ddots & \ddots & \vdots \\
\Gamma^{(\ell-1)} & \cdots & \cdots & \Gamma^{(0)}
\end{pmatrix},
\end{equation}
where 
\begin{equation}
    \Gamma^{(m)}=\frac{i}{L}\sum_k\tanh \frac{\mu_k}{2}  
    \begin{pmatrix}
        0 &g^{(m)}(k) \\
        -g^{(m)}(-k) &0
    \end{pmatrix},
\end{equation}
and
\begin{align}
    g^{(m)}(k)=   \left(u_k^* v_k+u_k v_k^*\right)\sin(m k)-\\
    (|u_k|^2-|v_k|^2)\cos(mk). \nonumber
\end{align}

\section{Mpemba effect at asymptotically long times}

\subsection{\new{Linearization of the Dissipator}}
In this paper we have used the GGE description of the time evolution of the density matrix, which reduces the number of parameters from exponentially many to linear in system size. The cost we pay is that the equation describing the time evolution $\dot{\lambda}_i(t)$ of the Lagrange parameters, Eq.~(\ref{eq:dlambda}), is nonlinear in Lagrange parameters. 

At late times, very close to the steady state, we can approximate the equations of motion with a linear operator by expanding them in first order in deviations of the Lagrange parameters $\delta\lambda_i$ from their steady state values $\lambda_{\infty, i}$, $\lambda_i = \lambda_{\infty, i} + \delta \lambda_i$. Using this expansion, the GGE state can be expressed to the first order in $\boldsymbol{\delta \lambda}$ as:
\begin{equation}
\label{eq:rho_taylor}
    \rho_{\boldsymbol\lambda} \approx \rho_{\boldsymbol{\delta\lambda}} = \rho_{\infty} \left(1-\sum_i^{N_C} \delta\lambda_i \left(C_i - \langle C_i \rangle_{\infty}\right)\right),
\end{equation}
where $\rho_{\infty}$ is the steady state given by  $\boldsymbol{\lambda_{\infty}}$, and $\langle \cdot \rangle_{\infty}$ denotes the expectation value of an operator in the steady state. The above expression is obtained using expansions:
\begin{align}
    e^{\sum_i^{N_C} \lambda_i C_i} 
     \approx e^{\sum_i^{N_C} \lambda_{\infty, i} C_i} \bigg(1-\sum_i^{N_C} \delta\lambda_i C_i \bigg), \notag  \\
    Z = \tr[e^{\sum_i^{N_C} \lambda_i C_i}] \approx Z_{\infty} \bigg(1-\sum_i^{N_C} \delta\lambda_i \langle C_i \rangle_{\infty} \bigg).
\end{align}

We insert the first order expansion of the density matrix (\ref{eq:rho_taylor}) into Eq.~(\ref{eq:dlambda}) of the main text to obtain:
\begin{align}
    &\dot{\lambda}_j = \dot{\delta\lambda}_j \notag \\
    &\approx-\sum_{k=1}^{N_C}\chi_{jk}^{-1}(\boldsymbol{\lambda}) \, \tr\Big[C_{k} \cl{D} \big(\rho_{\infty}-\rho_{\infty}\sum_{i=1} ^{N_C}\delta\lambda_i \left(C_i - \langle C_i \rangle_{\infty}\right)\big)\Big]  \notag \\
    &\approx\sum_{i=1}^{N_C} \sum_{k=1}^{N_C}\chi_{\infty, jk}^{-1} \, \tr \left[C_{k} \cl{D} \left(\rho_{\infty} C_i\right)\right]\delta\lambda_i.
\label{eq:dlambda_lin_derivation}
\end{align}
Here, only the zeroth order of the connected correlator $\chi_{\infty, jk}(t) = \langle C_j C_{k}\rangle_{\infty}-\langle C_j\rangle_{\infty}\langle C_k\rangle_{\infty}$ contributes to the first order expansion of the equation of motion.
Moving from the second to the third line, we used twice that the dissipator acting on the steady state produces no flow, namely $\tr[C_{k} \cl{D} \big(\rho_{\infty}\big)] = \tr[C_{k} \langle C_i \rangle_{\infty}\cl{D} \big(\rho_{\infty} \big)] = 0$.

The dissipator $\hat{\mathcal{D}}$ is proportional to the coupling strength $\epsilon$, which in the zeroth order approximation $\rho_{\boldsymbol{\lambda}}$ only plays the role of rescaling time. If we introduce a transformation of variables $\tau = \epsilon t$ into the equation of motion for Lagrange parameters, e.g. Eq. \eqref{eq:dlambda} or Eq. \eqref{eq:dlambda_lin_derivation}, equations obtained depend only on $\tau$ and not on $\epsilon$. To make notation cleaner, we use the above transformation of variables in the following sections. 

Now we can write the linear equations of motion for the Lagrange parameters close to the steady state:
\begin{align}
\label{eq:dlambda_lin}
    &\dot{\delta\lambda_j} (\tau) = \sum_{i=1}^{N_C} D_{ji} \,\delta\lambda_i (\tau) \notag  \\ &D_{ji} = \sum_{k=1}^{N_C}\chi_{\infty, jk}^{-1} \, \tr\big[C_{k} \cl{D} \left(\rho_{\infty} C_i\right)\big],
\end{align}
where the time derivative denoted with the dot $\dot{\delta\lambda_j} (\tau)$ is now taken over $\tau$.

One can compute the entries of the linearized dissipator as follows. First, the steady-state Lagrange parameters are determined either by solving the nonlinear fixed-point equation Eq.~\eqref{eq:dlambda} or by time evolution from initial conditions. Second, using these parameters, one evaluates the matrix elements of the linear operator according to the definition above. These steps can be carried out using exact diagonalization, tensor network methods for large systems, or a free-fermion representation in the non-interacting integrable case. For sufficiently large systems, the resulting entries of the linearized dissipator become independent of system size.

\begin{figure}
\centering
\includegraphics[width=\columnwidth]{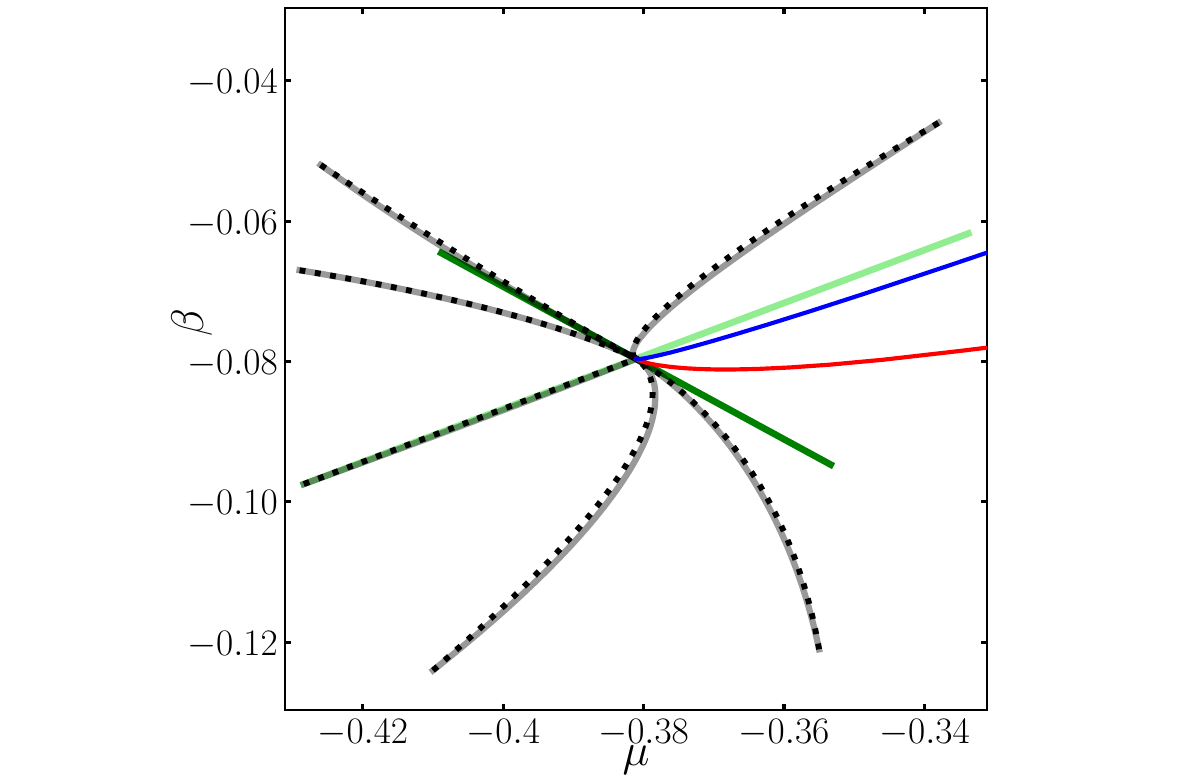}
\caption{Linearization of the equations of motion for Lagrange parameters for the model with $N_C = 2$ conserved quantities \eqref{eq:H2}. 
The light (fast direction) and the dark (slow direction) green lines represent the right eigenvectors of the linear operator $D$ in the space of the Lagrange parameters from Eq.~\eqref{eq:dlambda_lin}. The grey trajectories are obtained from the nonlinear coupled Eq.~\eqref{eq:dlambda}, while the black dotted ones are obtained from the linear approximation Eq. \eqref{eq:time_uk}. Blue and red curves correspond to trajectories considered in the main text. Calculated with tensor network on $L = 80$ sites.
}
\label{figure:S1}
\end{figure}

For a non-interacting integrable Hamiltonian, one can perform the linearization also on the level of the scattering Eqs.~\eqref{eq::nqdot} for momentum occupation evolution. We define deviations of momentum occupations from their steady state values $\delta\langle n_q \rangle = \langle n_q \rangle - \langle n_q \rangle_\infty$ and obtain the matrix governing the linearized dynamics $D^{(n)}$:
\begin{equation}\label{eq::ndot_lin}
\delta \langle \dot{n}_q \rangle = \sum_k D^{(n)}_{qk}\,\delta \langle n_k \rangle , \,\,\,\,\,
D^{(n)}_{qk} =
\left.
\frac{\partial \langle \dot {n}_q\rangle}{\partial \langle n_k\rangle}
\right|_{\langle n\rangle=\langle n\rangle_{\infty}}
\end{equation}
For our integrable setup the diagonal matrix elements result to
\begin{align}\label{eq::ndot_lin-1}
D^{(n)}_{qq}
=
\frac{2}{L}
\sum_{q'}
\Big[
&
-f^{s}_{q,q'} \langle n_{q'}\rangle_\infty
-
f^{s}_{q',q} \langle1- n_{q'}\rangle_\infty
\notag \\
&
-f^{c}_{q,q'}\langle1- n_{q'}\rangle_\infty
-
f^{a}_{q,q'} \langle n_{q'}\rangle_\infty
\Big],
\end{align}
and for \(k\neq q\),
\begin{align}\label{eq::ndot_lin-2}
D^{(n)}_{qk}
=
\frac{2}{L}
\Big[
&
f^{s}_{q,k} \langle 1- n_{q}\rangle_\infty
+
f^{s}_{k,q} \langle n_{q}\rangle_\infty
\notag \\
&
-f^{c}_{q,k}\langle1- n_{q}\rangle_\infty
-
f^{a}_{q,k} \langle n_{q}\rangle_\infty
\Big].
\end{align}

\subsection{\new{Relaxation in the space of the linear dissipator}}
Let us rewrite Eq.~\eqref{eq:dlambda_lin} as $\boldsymbol{\dot{\delta\lambda}} = D \, \boldsymbol{\delta\lambda}$, where bold symbols denote vectors and $\boldsymbol{\delta\lambda} = (\delta\lambda_1, \dots \delta\lambda_{N_C})^{T}$. We assume that the matrix $D$ is diagonalizable with nondegenerate eigenvalues $\alpha_k$ and corresponding right eigenvectors $r^{(k)}$, as is true for the example considered later. Let $V$ denote the matrix whose columns are the eigenvectors $r^{(k)}$, such that $V^{-1} D V = \text{diag}(\alpha_1, \dots \alpha_{N_C})$. Eq. \eqref{eq:dlambda_lin} can then be expressed in the eigenspace of the linearized dissipator as
\begin{equation}
    \label{eq:diag_dlambda}
    \mathbf{\dot{u}} = \text{diag}(\alpha_1, \dots \alpha_{N_C}) \, \mathbf{u}, \qquad \mathbf{u} = V^{-1} \boldsymbol{\delta\lambda}.
\end{equation}
Components of the trajectory in the eigenspace of the matrix $D$ independently decay exponentially
\begin{equation}
\label{eq:time_uk}
    u_k(\tau) = e^{\alpha_k \tau} u_k (0).
\end{equation}
We assume that at initial time $\tau = 0$ the trajectory $\mathbf{u}(0) = V^{-1} \boldsymbol{\delta\lambda}(0)$ is already in the linearized regime.
We sort the eigenmodes by their decaying real parts, $\mathrm{Re}(\alpha_{N_C}) < \dots < \mathrm{Re}(\alpha_1) < 0$, so that the largest eigenvalue $\alpha_1$ corresponds to the slowest decay mode, while the smallest $\alpha_{N_C}$ gives the fastest decay mode.

We use tensor networks on $L = 80$ sites to calculate the steady state parameters and construct the $2 \times 2$ matrix $D$ for the setup in which energy and magnetization are the only conserved quantities of the Hamiltonian, Eq.~\eqref{eq:H2} of the main text. Two right eigenvectors are obtained and plotted in Fig.~\ref{figure:S1} in the space of Lagrange parameters $(\beta, \mu)$, with the fast mode $r^{(2)}$ shown in light green and the slow mode $r^{(1)}$ in dark green. They correspond to real eigenvalues $\alpha_2 \approx -5.058$ and $\alpha_1 \approx -3.155$, and differ from the largest non-zero eigenvalues obtained from the diagonal ensemble on $L = 14$ sites (see End Matter) for less than $5\%$.
Trajectories for a few initial conditions are obtained in two ways: from nonlinear coupled Eqs.~\eqref{eq:dlambda} (gray lines), and from the linear approximation Eq.~\eqref{eq:time_uk} (black dotted lines). The gray and black trajectories nearly coincide, indicating that the linear approximation is accurate at the considered distances from the steady state. Notice from Eq. \eqref{eq:time_uk} that the trajectories always approach the steady state along the slow eigenvector.

\subsection{\new{Distance between two close-by GGEs}}
We use the first order expansion of the density matrix around the steady state Eq. \eqref{eq:rho_taylor} to derive the first order expansion of the normalized Frobenius distance, Eq.~\eqref{eq:d_norm}, in terms of $\boldsymbol{\delta\lambda}$. This can be done for any two close GGE states, but we write it here for the steady state and a state close to it:
\begin{align}\label{eq:d_norm_lin}
&d_{\ell}(\rho_{\boldsymbol{\delta\lambda}},\rho_{\infty})^2 
\approx d_{\ell}(\boldsymbol{\delta\lambda})^2
= \sum_{i,j}^{N_C} M_{ij}^{(\ell)} \, \delta\lambda_i\delta\lambda_j \\
&M_{ij}^{(\ell)} = \frac{\tr\big[\tr_{\bar{A}}[\rho_{\infty} \left(C_i - \langle C_i\rangle_{\infty} \right)] \ \tr_{\bar{A}}[\rho_{\infty} \left(C_j - \langle C_j\rangle_{\infty} \right)]\big]}{2\tr\big[ \ \tr_{\bar{A}}[\rho_{\infty}]^2 \ \big]}. \notag
\end{align}
Here, we perform the trace over the complement of subsystem $A$ to obtain the distance of reduced density matrices on  compact support $\ell=|A|$. Equidistant points lie on an ellipsoid in the GGE manifold. 

In the case of two conserved quantities, Eq. \eqref{eq:H2}, the distance is given by:
\begin{equation}\label{eq:d_norm_lin_Nc2}
d_{\ell}(\delta\beta, \delta\mu)^2
=  M_{\beta\beta}^{(\ell)} \,  \delta\beta^2 + M_{\mu\mu}^{(\ell)} \,  \delta\mu^2 + 2 M_{\beta\mu}^{(\ell)} \,  \delta\beta\delta\mu.
\end{equation}
Equidistant points lie on a rotated ellipse. Parameters $M_{ij}^{(\ell)}$ are given by the steady state values of Lagrange parameters and can be computed using exact diagonalization, tensor network methods, or free fermion methods, similar to the linearized dissipation operator in Eq.~\eqref{eq:dlambda_lin}.

\subsection{\new{Geometric condition for the Mpemba effect at late times}}

By expressing the distance to the steady-state \eqref{eq:d_norm_lin} in the eigenspace of the linearized dissipator \eqref{eq:diag_dlambda} and using Eq.~\eqref{eq:time_uk}, we obtain its time evolution:
\begin{align}
d_{\ell}(\boldsymbol{\delta\lambda}(\tau))^2 
&= \boldsymbol{\delta\lambda}(\tau)^T M^{(\ell)} \boldsymbol{\delta\lambda}(\tau) \notag \\
&= \boldsymbol{\delta\lambda}(\tau)^T (V^T)^{-1} V^{T} M^{(\ell)} V V^{-1} \boldsymbol{\delta\lambda}(\tau) \notag \\ 
&= \mathbf{u}(\tau)^{T} \tilde{M}^{(\ell)} \mathbf{u}(\tau) \notag \\
&= \sum_{k, l}^{N_C} \tilde{M}^{(\ell)}_{k, l}\,u_k(0)u_l(0) \,e^{(\alpha_k + \alpha_l)\tau}.
\end{align}
where $\tilde{M}^{(\ell)} = V^{T}M^{(\ell)}V$. For asymptotically long times, all of the modes will have decayed except for the slowest:
\begin{equation}\label{eq:lin_distance_long}
    d_{\ell}(\boldsymbol{\delta\lambda}(\tau \gg 1)) \sim \sqrt{\tilde{M}^{(\ell)}_{11}} \, |u_1(0)| \, e^{\alpha_1 \tau}.
\end{equation}
Therefore, \textit{at long times, our distance to the steady state is given by the initial value of the slowest component} $u_1(0)$. In fact, this is true for any reasonable proximity measure $\frak{D}$.

Using $u_1(0)$ as a universal proximity measure at long times, we identify a sufficient condition for the Mpemba effect to occur for the initial conditions in the linearized regime.
If the distance $\frak{D}$ to the steady state of initial state (1) is smaller than that of initial state (2), while the absolute value of the component along the slowest mode is larger for state (1) than for state (2), then the Mpemba effect occurs:
\begin{align}\label{eq:cond1}
\frak{D}(\boldsymbol{\delta\lambda}^{(1)}(0)) < \frak{D}(\boldsymbol{\delta\lambda}^{(2)}(0)), \,\,\,|u^{(1)}_1(0)| > |u^{(2)}_1(0)|.
\end{align}
The opposite situation also implies the Mpemba effect, 
\begin{align}\label{eq:cond2}
\frak{D}(\boldsymbol{\delta\lambda}^{(1)}(0)) > \frak{D}(\boldsymbol{\delta\lambda}^{(2)}(0)), \,\,\,
|u^{(1)}_1(0)| < |u^{(2)}_1(0)|.
\end{align}

\begin{figure}
\centering
\includegraphics[width=1\columnwidth]{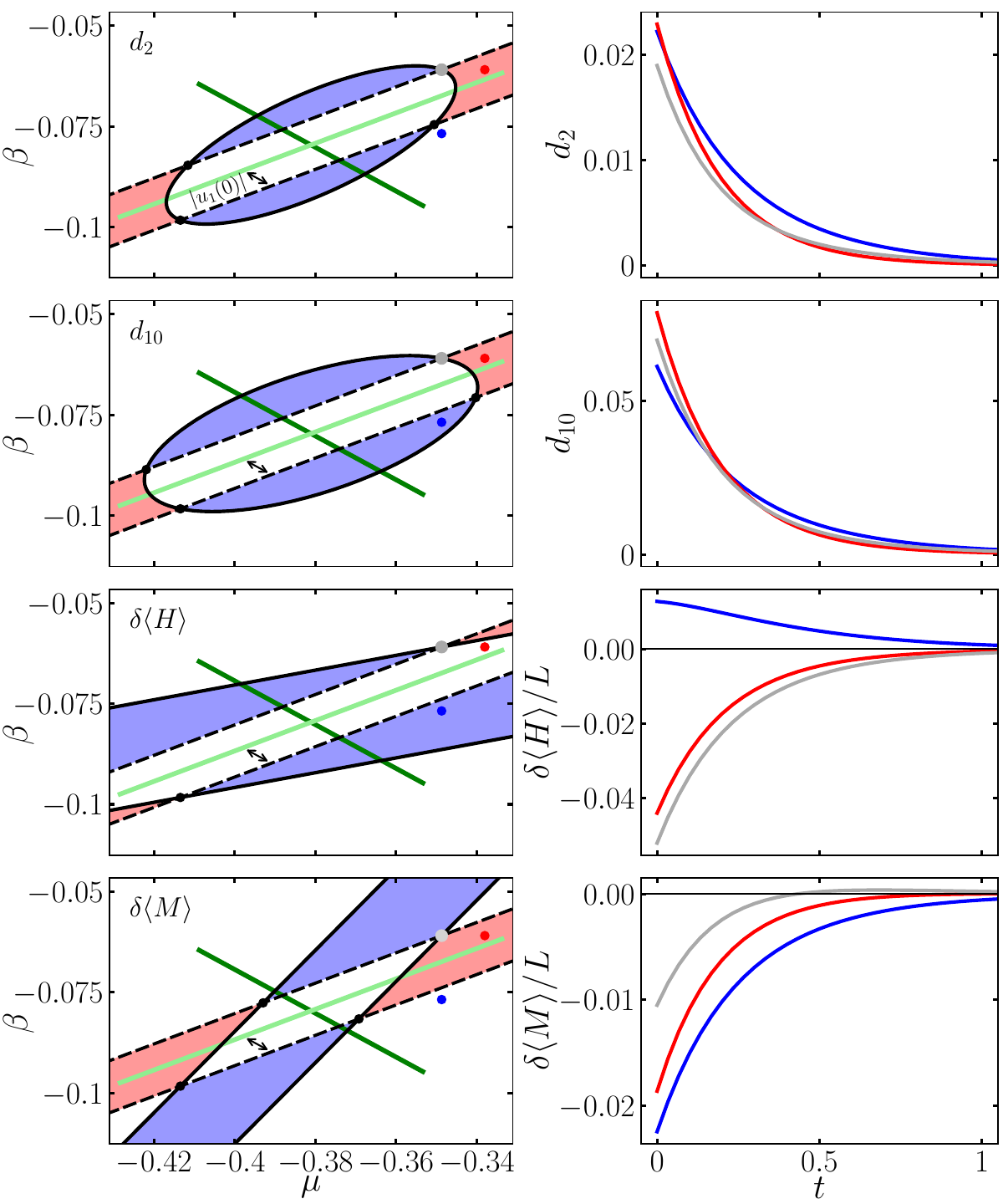}
\caption{
Left: Geometric condition for the occurance of the Mpemba effect in the linearized regime, Eqs. \eqref{eq:cond1} and \eqref{eq:cond2}.
The first row shows the condition for the normalized Frobenius distance on subsystem of 2 sites, $d_2$. The black ellipse represents states with the same distance $d_2$ to the steady state as the reference gray point, while the dashed lines are parallel to the fast eigenvector (light green) and represent states with the same initial value of the slow component $|u_1(0)|$ as the gray point. The blue and red region states exhibit the Mpemba effect relative to the gray point satisfying conditions \eqref{eq:cond1} (red) and \eqref{eq:cond2} (blue).
Equivalently, we show the regions for the distance on 10 sites $d_{10}$ (second row), the energy $\langle H \rangle$ (third row) and magnetization $\langle M \rangle$ (fourth row).
Right: Time evolution of the distances to the steady state and observable deviations from the steady state value for three initial states corresponding to the gray, blue and red points in the left column. Calculated with tensor network on $L = 80$ sites.
}
\label{figure:S2}
\end{figure}

In Fig. \ref{figure:S2} in the left column, we plot the condition geometrically in the space of Lagrange parameters for the case of two conserved quantities, Eq. \eqref{eq:H2}, and the Lindblad dissipator considered in the main text. In the first row, we show the condition for $\frak{D}=d_2$ and in the second for $\frak{D}=d_{10}$. For all of the cases, we choose an initial reference state (1) denoted with a gray point. All points with the same distance as the gray point are denoted with the black solid line. All points with the  same slow component $|u_1^{(1)}(0)|$ as the gray point are denoted with dashed lines. 
For each distance, we can identify the two regions of initial states (2), which will exhibit the Mpemba effect when compared to the gray point (1): 
the red region satisfies condition \eqref{eq:cond1} and the blue region satisfies condition \eqref{eq:cond2}.
The black dots denote the other three initial conditions, which have the same initial and asymptotical distance.

From the regions exhibiting the Mpemba effect, it is clear that the geometric condition depends on the choice of distance $\frak{D}$. This is illustrated explicitly in the right column where we plot time evolution starting from the red and blue initial points shown in the left column. Depending on the chosen distance, these states may or may not exhibit the Mpemba effect relative to the gray initial state. The difference between the first two rows originates from the distinct subsystem-size $\ell$ dependence of the entries $M^{(\ell)}_{\beta \beta}, M^{(\ell)}_{\beta \mu}$ and $M^{(\ell)}_{\mu \mu}$ that enter the distance in Eq. \eqref{eq:d_norm_lin}. For sufficiently large $\ell$ and $L$, however, all entries are expected to scale uniformly with $\ell^2$ \cite{fagotti13gge2}, implying that the geometric Mpemba condition becomes independent of $\ell$. The subsystem sizes for which this universal behavior is expected depend on the steady-state Lagrange parameters $\boldsymbol{\lambda}_{\infty}$ entering the definition of $M_{ij}^{(\ell)}$. 
For the present case with two conserved quantities, it is likely much larger than the system sizes accessible in our tensor network simulations.

However, for the non-interacting integrable model, where scattering equations allow us to study much larger subsystem sizes $\ell$, Fig.~\ref{figure:fig2}(d) from the main text already hints towards such universal behaviour and the corresponding saturation of the crossing time $\epsilon t_{Mp}=\tau_{Mp}$ with $\ell$. 
Namely, if we consider two trajectories whose normalized Frobenius distances become equal at the crossing time $\tau_{Mp}$ for a sufficiently large $\tilde{\ell}$:
\begin{equation}
    d^{(\tilde{\ell})} (\boldsymbol{\delta\lambda}^{(1)}(\tau_{Mp})) = d^{(\tilde{\ell})} (\boldsymbol{\delta\lambda}^{(2)}(\tau_{Mp})) = d^{(\tilde{\ell})}
\end{equation}
Then, for all $\ell > \tilde{\ell}$ there will be a crossing at the same time $\tau_{Mp}$, albeit at a different distance:
\begin{equation}
    d^{(\ell)} (\boldsymbol{\delta\lambda}^{(1)}(\tau_{Mp})) = d^{(\ell)} (\boldsymbol{\delta\lambda}^{(2)}(\tau_{Mp})) = \frac{\ell}{\tilde{\ell}} \, d^{(\tilde{\ell})}
\end{equation}
We checked that at the crossing times $\tau_{Mp}$ in Fig.~\ref{figure:fig2}(d), the quasiparticle occupations are sufficiently close to the steady state that the linearization and the above argument for the saturation of $\tau_{Mp}$ with $\ell$ can be used.

\subsection{\new{Mpemba effect and observables}}

One can also linearize the difference in the expectation value of an observable $O$ to the steady state value:
\begin{equation}\label{eq:obs_lin}
\tr[O \rho_{\boldsymbol{\delta\lambda}}] - \tr[O \rho_{\infty}]  \approx \delta \langle O (\boldsymbol{\delta\lambda}) \rangle = -\sum_{i}^{N_C} \langle O C_i \rangle_{\infty, c} \  \delta \lambda_i,
\end{equation}
where $\langle O C_i \rangle_{\infty, c}$ is the connected correlation function between the observable $O$ and the conserved quantity $C_i$ evaluated in the steady state. States with the same absolute ``distance'' of observable expectation value to its steady state value lie on two parallel hyperplanes in the GGE manifold, i.e., in the case of two conserved quantities on two parallel lines. In Fig.~\ref{figure:S2} (left column) we show them in solid lines with respect to the reference gray point for the energy (third row) and magnetization (fourth row) expectation values in model \eqref{eq:H2} with two conserved quantities and the dissipator considered in the main text. 

Similarly to Eq. \eqref{eq:lin_distance_long} for distances, one can express \eqref{eq:obs_lin} in the eigenspace of the linearized dissipator and conclude that the slow component of the initial state $u_1(0)$ determines the long-time dynamics of observables. The dashed lines in the left column of Fig.~\ref{figure:S2} indicate pairs of Lagrange parameters with the same slow-component value as the gray reference point.

Combining the above, initial states (2) exhibiting the Mpemba-like effect in observable expectation values \eqref{eq:obs_lin} with respect to the gray initial state (1) are shown again as red and blue regions, and satisfy the geometric conditions \eqref{eq:cond1} and \eqref{eq:cond2}, respectively. By a Mpemba-like effect, we mean a situation in which an initial state with an expectation value farther from the steady-state value eventually gets closer to it than another initial state. First, we find that the identified regions depend on the observable considered. This is highlighted in the right column showing time-dependent expectation values obtained by starting from three different initial states (gray, red and blue points); the Mpemba-like effect is observed for the gray and blue point when measuring energy, while it is observed for gray and red point when measuring magnetization. Secondly, identified Mpemba-like regions for observables do not overlap completely with the regions for the normalized Frobenius distance $\frak{D}=d_2, d_{10}$. That is not surprising - the expectation value of only one observable does not contain all information about the distance on the reduced density matrices of the same support, let alone on very large subsystems. The normalized Frobenius distance, on the other hand, quantifies all observables on supports smaller than or equal to the subsystem size. 

However, Eq.~\eqref{eq:d_norm_lin} does inspire us with ideas for an experimental reconstruction of distance $d_{\ell}$ without using more sophisticated tomographic techniques. Namely, if we can follow the time dependence of Lagrange parameters $\lambda_i(t)$
experimentally, we can easily reconstruct the ellipse parameters $M^{(\ell)}_{ij}$ via a classical numerical simulation using experimentally measured steady state $\boldsymbol{\lambda}_\infty$ as an input. $\lambda_i(t)$ and their steady state values can be obtained by measuring the nearly conserved quantities $\ave{C_i}(t)$ and performing a classical calculation. With such parametrization, we can experimentally reconstruct the distance behaviour,  Eq.~\eqref{eq:d_norm_lin}, with $\mathcal{O}(N_C)$ complexity, at least for cases where the Mpemba crossing happens close to the steady state. As we showed in the End Matter, Fig.~\ref{figure:figEM2}(b), equidistant points further away from the linearized regime lie on deformed ellipses, therefore the same trick formally cannot be applied for generic initial states far from the steady state, but it might nonetheless give some guidance and reduced complexity for experimental detection of the Mpemba effect.

\end{document}